\documentclass[12pt]{iopart}
\usepackage[dvips]{graphicx}
\usepackage{amssymb}
\usepackage{amsfonts}

\begin{document}

\title[Correlations in concentrated dendrimer solutions]
{Microscopic and coarse-grained correlation functions
in concentrated dendrimer 
solutions\footnote{This paper is dedicated to Professor Lothar Sch{\"a}fer 
on the occasion of his 60th birthday.}}

\author{I O G{\"o}tze and C N Likos}

\address{Institut f{\"u}r Theoretische Physik II,
Heinrich-Heine-Universit{\"a}t D{\"u}sseldorf,\\
Universit{\"a}tsstra{\ss}e 1, D-40225 D{\"u}sseldorf, Germany}

\begin{abstract} 
We employ monomer-resolved computer simulations of model
dendrimer molecules, to examine the significance of many-body
effects in concentrated solutions of the same. 
In particular, we measure the radial distribution functions and
the scattering functions between the centres of mass of the
dissolved dendrimers at various concentrations, reaching values
that slightly exceed the overlap density of the macromolecules. We 
analyse the role played by many-body effective
interactions by comparing the structural data to those obtained
by applying exclusively the previously obtained two-body effective
interactions between the dendrimers [G{\"o}tze~I~O, Harreis~H~M 
and Likos~C~N 2004 {\it J.~Chem.~Phys.} {\bf 120} 7761]. We find that the
effects of the many-body forces are small in general and they become
weaker as the dendrimer flexibility increases. Moreover, we test the
validity of the oft-used factorisation approximation to the total
scattering intensity into a product of the form- and the scattering
factors, finding a breakdown of this factorisation at 
high concentrations.
\end{abstract}

\section{Introduction}
\label{intro:sec}

Soft matter systems are characterised by the simultaneous existence
of two intrinsic structural length scales: the omnipresent atomic
or {\it microscopic} scale that is associated with the solvent molecules
and the monomers of the dissolved polymers (if any) and the 
{\it mesoscopic} scale that characterises the dissolved macromolecular
aggregates as a whole. Depending on the physical system under
consideration, the 
latter typically covers the range between several
nanometers and micrometers, spanning thereby three orders of
magnitude. The former is rather located in the domain of a few
{\AA}ngstr{\"o}m. In attempting to bridge the scales all the way from the
microscopic to the macroscopic ones, it has been proven very useful
to eliminate the atomic degrees of freedom from view, by performing
a statistical-mechanical trace over their degrees of freedom and
constructing thereby an {\it effective Hamiltonian} that involves
the mesoscopic degrees of freedom only \cite{likos:pr:01}. 
Although the effective Hamiltonian ${\mathcal H}_{\rm eff}$
greatly facilitates the
transition to the macroscopic scales, both its construction and
its interpretation have to be treated with care: indeed, the
effective potential energy function that involves the
mesoscopic degrees of freedom which appear  
in ${\mathcal H}_{\rm eff}$ are not true interaction potentials 
in the sense of Hamiltonian Mechanics but rather a constrained
free energy which arises by the thermodynamic trace of the microscopic
ones. 

There is a number of subtleties associated with the
effective potential energy function that have to be taken into
account when a coarse-grained statistical mechanical treatment
of a soft matter system is employed. Two of them are particularly
relevant in the context of calculating thermodynamic
quantities and tracing out phase diagrams. First, 
the potential energy cannot, in general, be written as a
sum of pair interactions:\footnote{An important
exception, however, is the depletion attraction in
colloid-polymer mixtures described by the idealised
Asakura-Oosawa model. In this case, all $n$-th order
polymer-mediated effective interactions between colloids
vanish identically for $n \geq 3$ if the polymer-to-colloid size ratio
does not exceed $2\sqrt{3}/3 - 1$. See Ref.\ \cite{brader:99}
for details.} the process of eliminating the microscopic
degrees of freedom inadvertently generates higher-order, many-body
potentials \cite{brader:99, dijkstra:prl:99, dijkstra:pre:99}. 
Truncating the effective potential energy function
at the pair-level constitutes the {\it pair potential approximation},
whose validity is not {\it a priori} guaranteed and
has to be explicitly checked.
And secondly, that the contributions to the potential
energy are in general state-dependent, the most prominent example
of the latter being the Debye-H{\"u}ckel effective pair potential
that has been extensively employed to model charge-stabilised
colloidal suspensions under certain physical conditions
\cite{lowen:hansen:00}. Sometimes the state-dependence of an effective
pair potential hides precisely the effect of many-body forces and
then particular care has to be taken in the ways in which the
pair potential is employed, so as to avoid blatant thermodynamic
inconsistencies \cite{ard:beware, stillinger1:03, stillinger2:03,
dijkstra:jcp:00}.

Many-body potentials are already encountered in the realm of atomic
systems, the Axilrod-Teller interaction \cite{axilrod:teller}
being a characteristic 
example that has been shown to be relevant for the description
of high-precision measurements of the structure factor of
rare gases \cite{tau:jpcm:99}. A formal decomposition of the
effective potential energy function between the particles of
one kind in a binary mixture in which the particles of the other
kind are traced out has been given in Refs.\ \cite{dijkstra:prl:99}
and \cite{dijkstra:pre:99}. Unfortunately, the treatment there 
applies only to mixtures for which the number densities of the
two components can be varied at will, e.g., colloid-polymer
or hard-sphere mixtures. It is not applicable to two broad
categories of soft matter systems, namely charged mixtures
and solutions of polymers of arbitrary architecture. In the
former case, the number densities of the two components are
constrained by the electroneutrality condition. In the latter,
where one specific monomer \cite{likos:prl:98, jusufi:jpcm:01} or the centre 
of mass of the molecule \cite{krueger:etal:98, ard:prl:00, ard:pre:00,
ard:pre:01, ard:jcp:01} are chosen as 
effective, mesoscopic coordinates, the total number of monomers 
and the number of effective particles are coupled to each other
through the constraint of keeping the number of monomers per macromolecule
fixed. 

In charge-stabilised colloidal suspensions, three-body
forces are generated by nonlinear counterion screening. Their
effects have been examined by density functional theory and
simulations \cite{lowen:jpcm:98} as well as by numerical solution
of the nonlinear-Poisson Boltzmann equation \cite{dobn:prl:04,
dobn:pre:04}. It has been found that the three-body forces in this
case are {\it attractive} \cite{lowen:jpcm:98, dobn:prl:04, dobn:pre:04},
a result confirmed by direct experimental measurements using
optical tweezers \cite{dobn:prl:04, dobn:pre:04}. As far as
polymeric systems are concerned, the triplet forces in star
polymer solutions have been analysed by theory and simulations
in Ref.\ \cite{ferber:epje:01}, where it was found that they play
a minor role for concentrations vastly exceeding the overlap
density. For linear chains, on the other hand, the many-body
forces appear to have a more pronounced effect, as witnessed by
the considerable state-dependence of the effective pair potential
that reproduces the correlation functions of concentrated polymer
solutions \cite{ard:pre:01, ard:jcp:01}. The general functional form of
the centre-of-mass effective interaction between polymer chains
was found to preserve its Gaussian form, its strength and range
being nevertheless modified within a range of $\sim 10\%$ of their
original values, due to many-body effects \cite{ard:pre:01, ard:jcp:01}. 

Another polymeric system that serves as a prototype for a tunable
colloidal system that displays a Gaussian, soft effective pair
interaction is that of a solution of dendritic macromolecules,
or dendrimers for simplicity \cite{likos:ac:04}. It has been 
recently shown that a Gaussian effective pair potential can describe
extremely well the scattering intensities obtained experimentally
from concentrated dendrimer solutions \cite{likos:macrom:01, likos:jcp:02}.
The Gaussian pair interaction has also been explicitly measured
in recent computer simulations that employed two different 
coarse-grained models for the microscopic, monomer-monomer
interactions \cite{ingo:jcp:04}. Nevertheless, in the approach
of Ref.\ \cite{ingo:jcp:04} only {\it two} dendritic molecules
were simulated, hence no information about many-body forces
was gained. In the present work, we address the issue of the
magnitude and importance of many-body effective interaction potentials
in concentrated dendrimer solutions. We do not attempt to derive an
explicit decomposition of the potential energy function into $n$-body
terms, $n = 2, 3, 4, \ldots$; this would require separate simulations
of just $n$ dendrimers. Instead, we explicitly simulate a large
number of interacting dendrimers at the microscopic level simultaneously.
We measure thereby the pair correlation functions in the concentrated
system directly and we compare the result with the one obtained 
by simulating the {\it same} number of dendrimers as effective
entities interacting exclusively by means of pair potentials. 
The discrepancies in the results from the two approaches for the
correlation functions yield then information regarding the importance
of the many-body forces {\it of all orders}. We find that the
many-body effects are of minor importance, especially for flexible
dendrimers. 

The rest of the paper is organised as follows. In section 
\ref{model:sec} we present our model and the simulation details.
In section \ref{pair:sec} we present our results for the
correlation functions derived by the two approaches mentioned
above and we discuss the magnitude and origin of their
discrepancies. In section \ref{scatter:sec} we turn our attention
to the issue of the interpretation of the total scattering
intensities from concentrated dendrimer solutions, and in 
particular to the question of the validity of the 
so-called factorisation
approximation of the latter as the product 
of the form- and the structure factor, discussing 
the limits of applicability of such an approach. Finally, in
section \ref{summary:sec} we summarise and conclude.

\section{The model and simulation details}
\label{model:sec}

In this work, we focus exclusively on dendrimers of the fourth
generation (G4).
We model the macromolecules at the monomer-level using a simplified
model that pictures every monomer as a hard sphere of diameter
$\sigma$. The bonding between the connected monomers is modeled
by flexible threads of maximum extension $\sigma(1 + \delta)$.
In detail, the potential between {\it disconnected} monomers is given by
\begin{eqnarray}
V_{\rm HS}(r)=\left\{\begin{array}{l@{\qquad}l}\infty & 
\mbox{for}\quad r/\sigma < 1 \\0 & \mbox{for}\quad r/\sigma > 1 \end{array} \right.
\end{eqnarray}
whereas {\it bonded} monomers interact via the potential
\begin{eqnarray}
V_{\rm bond}(r)=\left\{\begin{array}{l@{\qquad}l}\infty & 
\mbox{for}\quad r/\sigma<1 \\0 & \mbox{for}\quad 1 < r/\sigma < 1+\delta\\
\infty & \mbox{for}\quad r/\sigma > 1+\delta \end{array} \right.
\end{eqnarray}

The quantity $\delta > 0$ serves as a control parameter of the
dendrimer conformations, with small $\delta$-values resulting into
stiff dendrimers and large values into loose structures. This
bead-thread model was originally introduced by Sheng 
{\it et al.} \cite{sheng:macrom:02}, who kept a fixed value $\delta = 0.4$
and examined the scaling of the radius of gyration of the dendrimers
as a function of generation number and spacer length. The same
model has been employed in a previous work by us, in order to
systematically examine the evolution of the dendrimers' conformational
properties with the generation number $G$ \cite{ingo:macrom:03}.
By comparing the results for various values of the parameter
$\delta$ and by performing a further comparison with results
from a different model, we have 
shown that the conformational properties 
of single dendrimers are insensitive with respect to the 
details of the microscopic model. Moreover, this very 
simple, coarse-grained model reproduces the experimental 
scattering data for isolated dendrimers very well \cite{ingo:macrom:03}.
As we are only interested in static properties, we also 
allow `ghost chains', i.e., crossing of bonds occurring 
for $\delta \ge \sqrt{2} - 1 \approx 0.414$ are possible.
Monte Carlo simulations of this model are very fast, 
as there is no need to calculate energies; one only needs
to check for overlaps, and 
additionally whether the conditions 
of the maximal bond extension are fulfilled.
If one of these conditions is violated, the trial move 
is rejected in any case, so the search for further overlaps can be aborted.
Furthermore, due to the very short range of the hard
sphere interaction, neighbour lists are very effective.

The effective {\it pair} interaction potential between the centres of mass
of two G4-dendrimers has been determined with the help of
configuration-biased Monte Carlo simulations of this model
in Ref.\ \cite{ingo:jcp:04}.  
The strength of the interaction between dendrimers can 
be tuned by varying the number of generations or the  
parameter $\delta$. Denoting by $r$ the centre of mass
separation, the 
$\delta$-dependent effective pair potential $V_{\rm eff}^{(2)}(r;\delta)$
has been found to have a Gaussian form with small, additional
corrections. In particular, it can be fitted by the function:
\begin{equation}
\fl
\beta\,V_{\rm eff}^{(2)}(r;\delta) = \epsilon_0 \exp\left(-\frac{r^2}{\gamma_0}\right)
          +\epsilon_1 \exp\left[-\frac{(r-r_1)^2}{\gamma_1}\right]
          -\epsilon_2 \exp\left[-\frac{(r-r_2)^2}{\gamma_2}\right],
\label{potential:fit}
\end{equation}
where $\beta = (k_{\rm B}T)^{-1}$ with Boltzmann's constant 
$k_{\rm B}$ and the absolute temperature $T$;
the numerical values of the various fit parameters,
depending on the choice of $\delta$, are given in Table
\ref{TABparameters}. Note that the precise values
of the fit parameters are slightly different than those
given in Ref.\ \cite{ingo:jcp:04}, since there we employed
a more constrained fit by setting $\gamma_0 = 4R_{g,\infty}^2/3$,
with the radius of gyration $R_{g,\infty}$ of the dendrimers
at infinite dilution, and 
$\epsilon_2 = 0$. The gyration radius is also shown at the
last column of Table \ref{TABparameters}. 
Here, we considered G4-dendrimers with two different 
values, $\delta=0.1$ and $\delta = 2.0$, representing the
two extreme cases studied in Ref.\ \cite{ingo:jcp:04}.

\begin{table}
\caption{The numerical values of the fit parameters of the effective pair
potential between the centres of mass of two G4-dendrimers appearing
in Eq.\ (\ref{potential:fit}) for two different values of $\delta$.
At the last column the gyration radius $R_{g,\infty}$ at infinite
dilution is also shown.}
\begin{center}
\begin{tabular}{lccccccccr}\hline\hline\hline
$\delta$ & $\epsilon_0$ & $\gamma_0/\sigma^2$ & $\epsilon_1$ & 
$\gamma_1/\sigma^2$ &
$r_1/\sigma$ & $\epsilon_2$ & 
$\gamma_2/\sigma^2$ & $r_2/\sigma$ & $R_{g,\infty}/\sigma$ \\ \hline 
0.1 & 55.75 & 9.75 & 5.0 & 0.9  & 2.5 & 0.1 & 1.5 & 7.2 & 2.665 \\
2.0 & 11.35 & 33.0 & 0.8 & 10.0 & 3.7 & 0.0 & --- & --- & 4.939 \\
\hline\hline\hline
\end{tabular}
\end{center}
\label{TABparameters}
\end{table}

Let $\rho = N/\Omega$ be the number density of a sample 
containing $N$ dendrimers enclosed in the volume $\Omega$.
The definition of the overlap density $\rho_{*}$ of a dendrimer
solution requires some care, as it is not a sharply defined
quantity. Previous simulation studies with this system 
\cite{ingo:macrom:03} have
revealed that the monomer density profiles around the dendrimer's
centre of mass decay to zero at a distance $r_{\rm c} \cong 1.5\,R_{g,\infty}$.
Motivated by this fact, we envision every dendrimer as a 
`soft sphere' of radius $r_c$ and define the overlap density
through the relation:\footnote{In the literature, 
there are alternative definitions.
For polymer chains, for instance, the definition 
$\frac{4\pi}{3}\rho_{*}R_g^3 = 1$ was used in Ref.\ \cite{ard:pre:01}.}
\begin{equation}
\frac{4\pi}{3}\rho_{*}r_{\rm c}^3 = 1.
\label{rhoov:eq}
\end{equation} 
Moreover, we introduce the diameter of gyration 
at infinite dilution, $\tau \equiv 2R_{g,\infty}$, as the characteristic
mesoscopic length scale to be used to 
introduce a dimensionless expression for the number density,
$\rho\tau^3$. In these terms, the overlap density of 
Eq.\ (\ref{rhoov:eq}) above is given by $\rho_{*}\tau^3 = 0.566$. 
The highest density in the simulation was 
$\rho_{\rm max}\tau^3 = 0.605$,
slightly exceeding the overlap value, since
$\rho_{\rm max} = 1.07\rho_{*}$.

For both 
values of $\delta$, ten different concentrations were
simulated, in particular at the densities
$\rho/\rho_{\rm max} = 0.1, 0.2, \ldots, 1.0$. 
Periodic boundary conditions were employed throughout.
At all densities, systems of 500 dendrimers were simulated, 
whereby each dendrimer consists of $\nu = 62$ monomers, and the size 
of the simulation box was changed in order to modify the
dendrimer number density.  
The minimum box length was 
$L_{\rm min} = 9.384\,\tau$, yielding a system with the 
density $\rho_{\rm max}$. The equilibration criterion for
the system at hand requires some care, as there are is no
internal energy in the microscopic model, since all interactions
are either zero or infinity. We therefore took advantage of the
fact that the {\it effective}, pair interaction $V_{\rm eff}^{(2)}(r;\delta)$
between the centres of mass is known and given by 
Eq.\ (\ref{potential:fit}) with the parameters given in 
Table \ref{TABparameters}. 
Hence, we chose to monitor the total effective pair potential
energy $U^{(2)}(N;\delta)$ given by
\begin{equation} 
U^{(2)}(N;\delta) = \frac{1}{2}\sum_{i=1}^{N}\sum_{j\ne i}^{N}
V_{\rm eff}^{(2)}(|{\bf r}_i - {\bf r}_j|;\delta),
\label{totalpairenergy:eq}
\end{equation}
where ${\bf r}_{i,j}$ denotes the position of the $i,j$-th 
centre of mass.

Two different starting configurations
were tried. In the first one,
the centres of mass dendrimers possessing identical microscopic
conformations were placed 
at the vertices of a fcc-lattice, which was achieved without
violation of the excluded volume conditions.
This procedure is particularly useful especially at the highest
density, $\rho_{\rm max}$, where a random distribution of the
centres of mass will result with high probability into 
a forbidden state with monomer overlaps. The system was
then equilibrated, monitoring $U^{(2)}(N;\delta)$ described above. 
In the second one, the
dendrimers' centres of mass were placed in a random arrangement.
Although this procedure requires a large number of failed
attempts before an allowed configuration is found, especially
at high densities, such configurations are possible. Once again,
we monitored the total effective pair potential energy during the
equilibration period, finding that it converges to the same
value as the one obtained from the fcc-initial state. In this
way, sufficient equilibration of the system was guaranteed. 
Finite-size effects were checked by selectively simulating 
some systems with 256 of dendrimers, in a box having
a correspondigly smaller volume,
so that the same density is achieved, and finding agreement between the
two attempts.

For $\delta=0.1$, $N_{\rm equil} = 10^7$ MC steps were used to 
equilibrate the system, and about 
$N_{\rm run} = 2 \times 10^8$ steps to gather statistics.
Statistical averages were calculated every $N_{\rm meas} = 10\,000$ MC steps.
For $\delta=2.0$, where a much larger random 
displacement for the monomers can be used, 
the equilibration phase 
consisted of $N_{\rm equil} = 10^6$ steps and statistical averages 
were calculated every $N_{\rm meas} = 1000$ steps over a period of 
$N_{\rm run} = 2 \times 10^8$ steps. The quantities measured
were monomer profiles around the centres of mass, the radial
distributions functions of the latter, 
radii of gyration, form factors, 
structure factors from the centres of mass
and total scattering intensities;
all these quantities will be
precisely defined in the sections that follow.

\begin{figure}
  \begin{center}
  \includegraphics[width=8.5cm,clip]{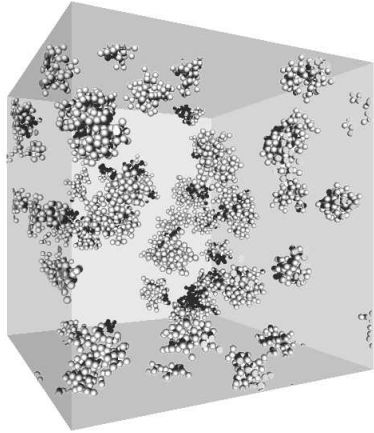}
  \end{center}
  \caption{A snapshot from the monomer resolved-simulation 
  of dendrimers. The monomers are rendered as spheres of diameter $\sigma$.
  Here, dendrimers with threads characterised through $\delta=0.1$
  at a density $\rho\tau^3 = 0.0605$ are shown. Note that only a part
  of the simulation box is shown,
  which has the same size as the full box depicted in Fig.\ \ref{rho10:fig}.}  
  \label{rho01:fig} 
\end{figure}
\begin{figure}
  \begin{center}
  \includegraphics[width=8.5cm,clip]{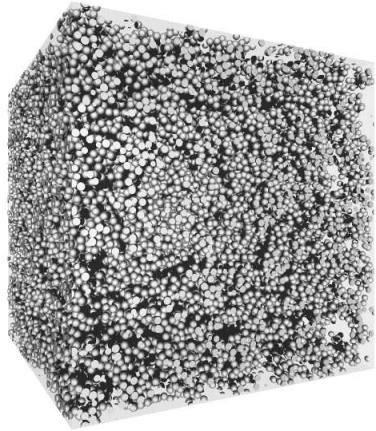}
  \end{center}
  \caption{Same as Fig.\ \ref{rho01:fig} 
  but at density $\rho\tau^3 = 0.605$. 
  Here the complete simulation box is shown.}
  \label{rho10:fig}
\end{figure}

In Figs.\ \ref{rho01:fig} and \ref{rho10:fig}, we show
simulation snapshots of the monomer-resolved simulations
for the lowest and the highest density for the thread length $\delta=0.1$.
(For clarity, in Fig.\ \ref{rho01:fig}
we show only a section of the simulation box of the same size
as in Fig.\ \ref{rho10:fig}.) Although at Fig.\ \ref{rho01:fig}
individual dendrimer molecules can still be resolved, since
the density is much smaller than $\rho_{*}$, in Fig.\ \ref{rho10:fig}
this is not any more possible. Here, $\rho = 1.07\rho_{*}$ and
the whole system appears as a dense solution of monomers, in which
the individual character of each macromolecule is lost. 
We will return to the implications of this fact in
section \ref{scatter:sec}.

In addition, a different kind of Monte Carlo simulations was
also carried out, in which the monomers were not explicitly
resolved. Instead, the dendrimers were replaced entirely by
their centres of mass, which were then treated as effective,
soft particles interacting exclusively by means of the
pair potential of Eq.\ (\ref{potential:fit}). 
Accordingly, we call this approach an {\it effective}
simulation. As all monomers
have dropped out of sight in the effective approach, it is only
possible to measure quantities pertaining to the centres of mass,
i.e., their radial distribution functions and structure factors.
Comparison of the results regarding these quantities that
are obtained through the two different types of simulations
yields important information by way of testing whether the
pair-potential approximation is meaningful.

\section{Comparison between the monomer-resolved and the effective
simulations}
\label{pair:sec}

Each dendrimer of the fourth generation consists of $\nu = 62$
monomers. Let $\alpha$, $\beta$ be monomer indices within
a given dendrimer whereas $i$, $j$ are integers describing the
dendrimer molecules as whole entities. In particular, let 
${\bf r}_i$ stand for the position of 
the centre of mass of the $i$-th dendrimer,
${\bf R}_{\alpha}^{i}$
denote the position vector of the $\alpha$-th monomer in
the $i$-th dendrimer, and ${\bf u}_{\alpha}^{i}$ stand for
the same quantity but now measured in a coordinate system
centred at ${\bf r}_i$. Obviously, it holds
\begin{equation}
{\bf R}_{\alpha}^{i} = {\bf r}_i + {\bf u}_{\alpha}^{i}.
\label{trafo:eq}
\end{equation}

In the monomer-resolved simulation, the following quantities
were measured: The radial distribution function $g(r)$ between the
centres of mass, defined as
\begin{equation}
g(r) = \frac{1}{N}\left\langle \sum_{i=1}^{N}\sum_{j \ne i}^{N}
       \delta\left({\bf r}-{\bf r}_{ij}\right)\right\rangle,
\label{gofr:eq}
\end{equation}
where $\langle \cdots \rangle$ denotes a statistical average and
${\bf r}_{ij} = {\bf r}_i - {\bf r}_j$. Related to this quantity
is the structure factor $S(q)$ that describes the correlations
between the centres of mass in reciprocal space and it is given by
\begin{equation}
S(q) = \frac{1}{N}\left\langle \sum_{i=1}^{N}\sum_{j=1}^{N}
       \exp\left[-{\rm i}{\bf q}\cdot
           \left({\bf r}_i - {\bf r}_j\right)\right]
       \right\rangle.
\label{sofq:eq}
\end{equation}
Note that $S(q)$ and $g(r)$ are related by a Fourier transformation
\cite{hansen:mcdonald}
\begin{equation}
S(q) = 1 + \rho\int{\rm d}^3r \exp\left[-{\rm i}{\bf q}\cdot{\bf r}\right]
           \left[g(r) - 1\right]. 
\label{ft:eq}
\end{equation}
Moreover, we took advantage of the microscopic nature of the simulation
to measure the dendrimers' form factor $F(q)$ at every simulated
density $\rho$. This quantity is expressed by the relation:
\begin{equation}
F(q) = \frac{1}{N}\sum_{i=1}^{N}
               \frac{1}{\nu}\left\langle \sum_{\alpha=1}^{\nu}
                                 \sum_{\beta=1}^{\nu}
               \exp\left[-{\rm i}{\bf q}\cdot
               \left({\bf u}_{\alpha}^{i} - {\bf u}_{\beta}^{i}\right)
               \right]\right\rangle,
\label{fofq:eq}
\end{equation}
Another quantity of interest is the monomer distribution around
the centre of mass, $\xi(u)$, which can again be measured at any desired
overall density and is given by the expression:
\begin{equation}
\xi(u) = \frac{1}{N}\sum_{i=1}^{N}
         \left\langle\sum_{\alpha = 1}^{\nu}\delta
         \left({\bf u} - {\bf u}_{\alpha}^{i}\right)\right\rangle,
\label{xiofr:eq}
\end{equation}
The overall size of the dendrimer is characterised by its radius
of gyration $R_g$, which was measured in the simulation by
calculating the quantity:
\begin{equation}
R_g = \frac{1}{N}\sum_{i=1}^{N}
      \sqrt{\frac{1}{\nu}\left\langle
      \sum_{\alpha=1}^{\nu}{\bf u}_{\alpha}^{i}\cdot{\bf u}_{\alpha}^{i}
      \right\rangle},
\label{rg:eq}
\end{equation}
In Eqs.\ (\ref{fofq:eq}) - (\ref{rg:eq}) above, the summand in the
sum over $i$ is the corresponding quantity (form factor, density
profile, and radius of gyration, respectively) of the $i$-th dendrimer.
The additional summation over $i$ and the division by the total 
number of dendrimers corresponds to an additional average over
{\it all} dendrimers. Since all macromolecules are equivalent,
the expectation values are identical for every summand. 
Finally, we also measured the scattering function $I(q)$
of the concentrated solution,
which corresponds to the coherent contribution of the total
scattering intensity in a SANS experiment, under the assumption
that all monomers possess the same scattering length density
\cite{mb:macrom:99, mb:mcp:00, mb:mcp:02, benoit}. 
This is given by the equation: 
\begin{equation}
I(q) = \frac{1}{N\nu}\left\langle
                     \sum_{i=1}^{N}\sum_{j=1}^{N}
                     \sum_{\alpha = 1}^{\nu}\sum_{\beta=1}^{\nu}
                     \exp\left[-{\rm i}{\bf q}\cdot
                         \left({\bf R}_{\alpha}^{i}-{\bf R}_{\beta}^{j}\right)
                         \right]\right\rangle,
\label{iofq:eq}
\end{equation}
i.e., it is the total coherent scattering intensity from all monomers of
the system.

In the effective picture, all information regarding the monomers'
degrees of freedom is lost, hence in the effective simulation we
can only measure the corresponding radial distribution function
$g_{\rm eff}(r)$ and the structure factor $S_{\rm eff}(q)$
of the centres of mass. These
are given by Eqs.\ (\ref{gofr:eq}) and (\ref{sofq:eq}) above
but with the averages now performed with the effective Hamiltonian,
i.e.,
\begin{equation}
g_{\rm eff}(r) = \frac{1}{N}\left\langle \sum_{i=1}^{N}\sum_{j \ne i}^{N}
     \delta\left({\bf r}-{\bf r}_{ij}\right)
     \right\rangle_{{\mathcal H}_{\rm eff}},
\label{gofreff:eq}
\end{equation}
and
\begin{equation}
S_{\rm eff}(q) = \frac{1}{N}\left\langle \sum_{i=1}^{N}\sum_{j=1}^{N}
       \exp\left[-{\rm i}{\bf q}\cdot
           \left({\bf r}_i - {\bf r}_j\right)\right]
       \right\rangle_{{\mathcal H}_{\rm eff}}.
\label{sofqeff:eq}
\end{equation}
The effective Hamiltonian ${\mathcal H}_{\rm eff}$ involves the momenta
${\bf p}_i$ and positions ${\bf r}_i$ of the centres of mass only
and contains exclusively pair interactions, i.e., 
\begin{equation}
{\mathcal H}_{\rm eff} = \sum_{i=1}^{N}\frac{{\bf p}_i^2}{2 m}
                         +\frac{1}{2}\sum_{i=1}^{N}\sum_{j\ne i}^{N}
                         V_{\rm eff}^{(2)}(|{\bf r}_i - {\bf r}_j|;\delta),    
\label{heff:eq}
\end{equation}
where $m$ is the dendrimers' mass, which is irrelevant as far
as static quantities of the system are concerned.
A particular property of the effective description of
a complex system is that it leaves all correlation functions
between the coarse-grained degrees of freedom invariant
{\it provided} that the mapping into the effective
system is {\it exact} \cite{likos:pr:01}. In other words,
if the effective Hamiltonian contains the contributions to
the effective potential at {\it all orders}, it makes no
difference whether one calculates quantities such as $g(r)$
or $S(q)$ in the original, microscopic description or in the
coarse-grained one. As our effective Hamiltonian ${\mathcal H}_{\rm eff}$
is truncated at the pair level, the deviations between 
$g(r)$ and $g_{\rm eff}(r)$ or, equivalently, between
$S(q)$ and $S_{\rm eff}(q)$ will be a measure of the importance
of the neglected many-body terms in Eq.\ (\ref{heff:eq}).

\begin{figure}
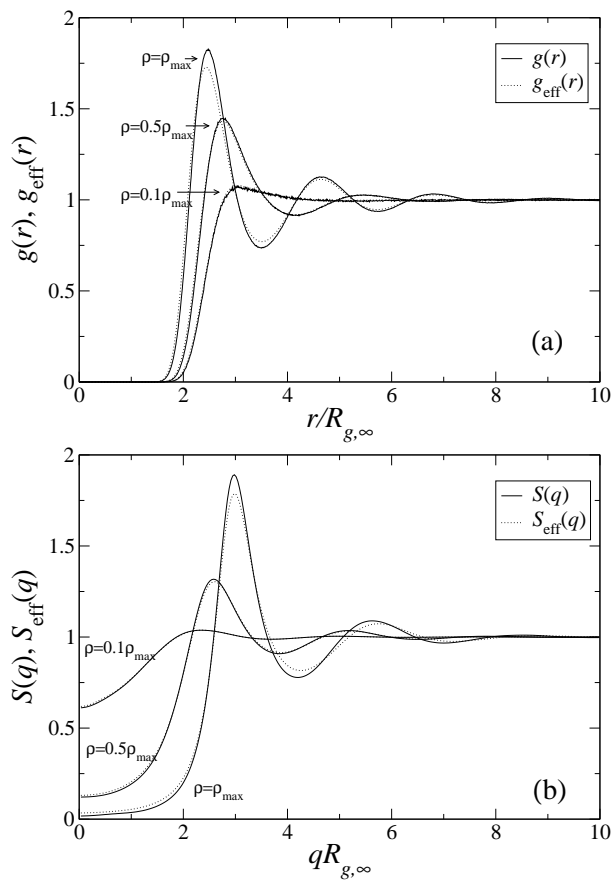

  \begin{center}
  \begin{minipage}[h]{7.2cm}
  \includegraphics[width=8cm,clip]{figure03a.eps}
  \includegraphics[width=8cm,clip]{figure03b.eps}
  \end{minipage}
  \end{center}
  \caption{Comparison between the results from the monomer-resolved
  and the effective simulation of concentrated dendrimers with maximal
  thread length $\delta = 0.1$ of the bonds. The three different
  densities are $\rho = 0.1\rho_{\rm max}$, $0.5\rho_{\rm max}$
  and $\rho_{\rm max}$, as indicated on the plots,
  with $\rho_{\rm max}\tau^3=0.605$. Results are shown for
  (a) the radial distribution function $g(r)$ and
  (b) the structure factor $S(q)$ of the centre of mass coordinates.}
  \label{G4_0.1_GrSq}
\end{figure}

\begin{figure}
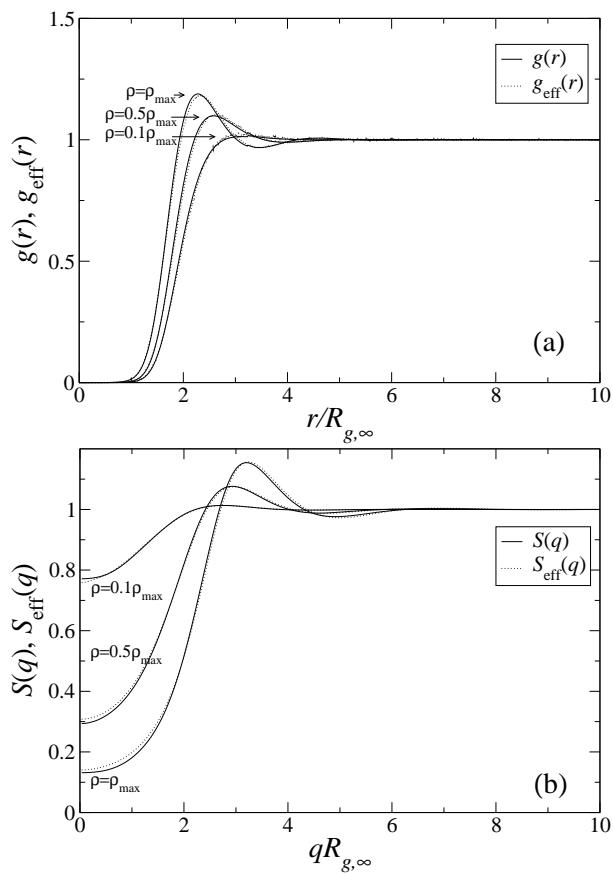

  \begin{center}
  \begin{minipage}[h]{7.2cm}
  \includegraphics[width=8cm,clip]{figure04a.eps}
  \includegraphics[width=8cm,clip]{figure04b.eps}
  \end{minipage}
  \end{center}
  \caption{Same as Fig.\ \ref{G4_0.1_GrSq}
           but for thread length $\delta=2.0$.}
  \label{G4_2.0_GrSq}
\end{figure}

Representative results comparing between the two approaches
are shown in Fig.\ \ref{G4_0.1_GrSq}, pertaining to 
the dendrimers with $\delta = 0.1$ and in Fig.\ \ref{G4_2.0_GrSq},
which refers to dendrimers with $\delta = 2.0$.
The length scale used in this plot is the zero-density
gyration radius of the dendrimers, $R_{g,\infty}$.  
For clarity, only the results only for three different densities 
obtained from the monomer resolved simulations are compared 
to those from the effective ones.  
At sufficiently low densities, $\rho = 0.1\rho_{\rm max}$,
the results from the two types of simulations are indistinguishable,
hence the pair-potential approximation is an excellent one
and many-body forces seem to play no role there; they can
be thus safely ignored. Deviations between the two descriptions
arise nevertheless as the overall concentration of the solution
grows. Referring to Fig.\ \ref{G4_0.1_GrSq}(a), we see that
for the $\delta = 0.1$-dendrimers, which have a rather high
internal monomer density, the deviations are already visible
(but small) at a density $\rho = 0.5\rho_{\rm max}$ and they
become more pronounced at the highest simulated density, $\rho = \rho_{\rm max}$.
The true radial distribution function $g(r)$ between the centres of
mass shows a more pronounced coordination than the effective one,
$g_{\rm eff}(r)$, and this effect is also reflected in the corresponding
structure factors. The peak height of $S(q)$ is higher than the one
of $S_{\rm eff}(q)$, pointing to the fact that the zero-density
pair potential underestimates somehow the strength of the repulsions
between the dendrimers' centres of mass. The relative deviation between the
two descriptions as far as the peak height
is concerned are at the highest density about $6\%$. Much more
drastic is the discrepancy of the $S(q \to 0)$ limit, for which
$S(q \to 0) = 0.018$ whereas $S_{\rm eff}(q \to 0) = 0.033$.
Given the fact that the $S(q = 0)$-value is proportional to the
osmotic isothermal compressibility of the solution, employing
the effective picture can lead here to serious errors in the
calculation of the thermodynamics of the system. Two integrations
of the inverse compressibility are needed in order to obtain the
Helmholtz free energy of the solution, hence errors at all lower
densities accumulate in performing such an integration and they
can lead to a serious underestimation of the free energy if the
effective picture is employed.  

The agreement between the microscopic and the coarse-grained
approaches is a lot better for the case of the $\delta = 2.0$-dendrimers,
which possess a much lower internal monomer density than their
$\delta = 0.1$-counterparts. Indeed, as can be seen in
Fig.\ \ref{G4_2.0_GrSq}(a), the radial distribution functions
$g(r)$ and $g_{\rm eff}(r)$ barely show any difference, all the
way up to the maximum density $\rho_{\rm max}$. Similar to the
case $\delta = 0.1$, $g(r)$ shows a slightly more pronounced 
coordination than $g_{\rm eff}(r)$, the difference 
between the two is nevertheless
extremely small. The same holds for the structure factors 
$S(q)$ and $S_{\rm eff}(q)$, shown in Fig.\ \ref{G4_2.0_GrSq}(b).
Here, even the discrepancy in the compressibility is very small,
with $S(q \to 0) = 0.132$ and $S_{\rm eff}(q \to 0) = 0.138$
at $\rho = \rho_{\rm max}$. For dendrimers with a higher degree
of internal freedom, the pair potential approximation holds 
all the way up to the overlap concentration. In this respect,
it is very satisfactory that it is precisely the model with 
the value $\delta = 2.0$ that has been found to accurately 
describe scattering data from real dendrimers \cite{ingo:jcp:04}.

\begin{figure}
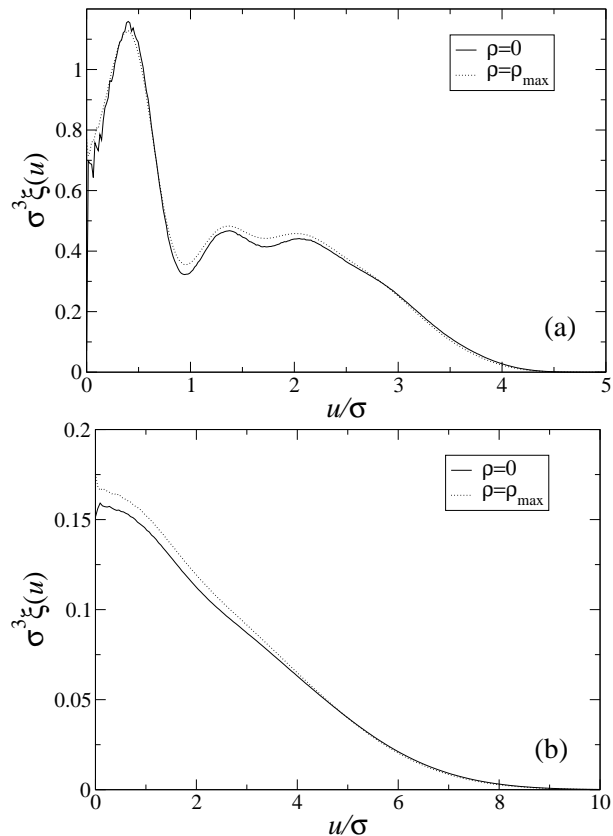

  \begin{center}
  \begin{minipage}[h]{7.2cm}
  \includegraphics[width=8cm,clip]{figure05a.eps}
  \includegraphics[width=8cm,clip]{figure05b.eps}
  \end{minipage}
  \end{center}
  \caption{The 
   radial monomer density profiles $\xi(u)$ [Eq.\ (\ref{xiofr:eq})]
   of the dendrimers
   around their centres of mass at infinite dilution ($\rho = 0$)
   and at the highest density $\rho = \rho_{\rm max} = 1.07\rho_{*}$,
   as indicated on the plots. (a) For model dendrimers with 
   thread length $\delta = 0.1$ and (b) for $\delta = 2.0$.
   Note the shrinkage and growth of the profiles.}
  \label{G4_rho}
\end{figure}

Let us now try to to obtain some physical insight into the mechanisms
that cause the true correlation functions to show higher ordering
than the effective ones. Suppose that the reason lied in the
increasing significance of three-body effective forces. 
Three-body potentials arise through three-dendrimer overlaps:
the region of space in which three spherical objects simultaneously
overlap is overcounted when one adds over the three pair interactions
and it has to be subtracted anew. Given the fact that any 
overlap between repulsive monomers gives rise to a correspondingly
repulsive interaction, together with the fact that the contribution
from the triple-overlap region has to be {\it subtracted}, leads
to the conclusion that triple forces should be {\it attractive},
as for the case of star polymers \cite{ferber:epje:01}, 
as well as self-avoiding polymer chains \cite{ard:pre:01}, for which
three-body forces have been measured explicitly \cite{foot:charge}.
Yet, an attractive contribution to the potential energy leads
to a {\it reduced} effective pair repulsion. This is 
on the one hand intuitively clear and, on the other, it can be
put in formal terms by making a density expansion of the density-dependent
pair interaction up to linear order in density, see Eq.\ (10) of
Ref.\ \cite{ard:pre:01}. Thus, we would then obtain   
a {\it weakening} of the correlations and an {\it increase}
of the osmotic compressibility, whereas in Figs.\ \ref{G4_0.1_GrSq} and
\ref{G4_2.0_GrSq} exactly the opposite is true. In order to obtain the
true $g(r)$ at $\rho = \rho_{\rm max}$ for the $\delta = 0.1$-dendrimers,
a renormalised effective pair potential $\tilde{V}_{\rm eff}^{(2)}(r;\delta,\rho)$
can be employed that is more strongly repulsive than the original
one, $V_{\rm eff}^{(2)}(r;\delta)$; as a matter of fact, we were 
able to reproduce $g(r)$ at $\rho_{\max}$ by using
$\tilde{V}_{\rm eff}^{(2)}(r;\delta = 0.1,\rho_{\rm max}) 
\cong 1.2\,V_{\rm eff}^{(2)}(r;\delta = 0.1)$.
A similar effect has been observed for polymer chains \cite{ard:pre:01}, 
for which the 
density-dependent, renormalised pair potential necessary to reproduce
$g(r)$ at high concentrations was found to be more repulsive than the
one that holds at $\rho = 0$, whereas, at the same time, the correction
arising from triplet forces alone goes in the opposite direction of
weakening the pair repulsions.

\begin{figure}
   \begin{center}
   \includegraphics[width=8cm,clip]{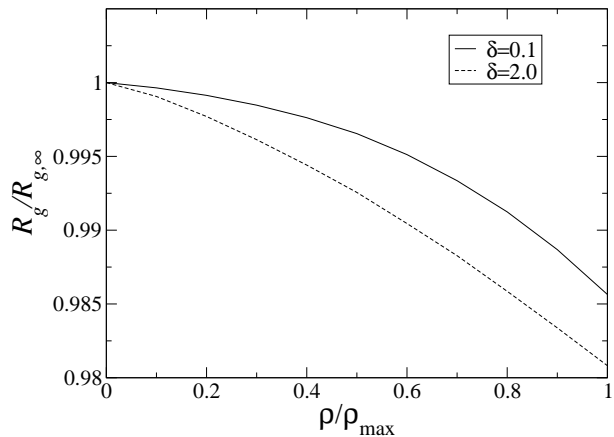}
   \end{center}
  \caption{The dependence of the dendrimers' radius of gyration
  on the solution density for the two types of model macromolecules,
  as indicated in the legend.}
  \label{Rg}
\end{figure}

The above considerations point to the fact that the
deviations between $g(r)$ and $g_{\rm eff}(r)$ are a genuinely 
many-body effect that arises from the high concentration of
the solution per se and cannot be attributed to three-body 
forces alone. In particular, the presence of many dendrimers
surrounding a given one in the concentrated solution, gives rise
to a deformation
of the dendrimer itself. To corroborate this statement, we
have measured the concentration-dependent monomer density profiles
$\xi(u)$ around the dendrimers' centre of mass, given by Eq.\ (\ref{xiofr:eq}).
Results are shown in Fig.\ \ref{G4_rho}(a) for the case $\delta = 0.1$
and in Fig.\ \ref{G4_rho}(b) for the case $\delta = 2.0$.
It can be seen that as a result of the crowding of the dendrimers
at the highest concentration, the monomer profiles become slightly
shorter in range and they grow in height; in other words, the dendrimers
{\it shrink} as a result of the increased overall concentration,
as can be also witnessed by the reduction of their radius of gyration
shown in Fig.\ \ref{Rg}. The molecules that effectively interact
are {\it stiffer} at higher densities than at lower ones; their 
internal monomer concentration grows with $\rho$ and as a result
of this deformation, the interaction between two
dendrimers becomes more repulsive than at zero density.

The above claim is supported by the fact that the effect of the 
concentration on the pair interaction is much more pronounced for
the dendrimers with the short thread length than for those with
the longer one. Although the monomer profiles for {\it both}
dendrimer kinds 
grow with $\rho$, the internal monomer concentration for the 
stiffer dendrimers is much higher than the one for the softer
ones. A concentration-induced increase of $\xi(u)$ has a much
stronger effect for the effective interaction of the stiff
dendrimers than for the soft ones, since it occurs at a scale
of $\sigma^3\xi(u) \sim 0.4$ for the former but at a scale of 
$\sigma^3\xi(u) \sim 0.1$ for the latter, see Fig.\ \ref{G4_rho}.
The monomer beads are modeled here as hard spheres.
The change in the
free energy of a hard-sphere fluid upon an increase
of the local density is highly nonlinear and grows rapidly
with increasing packing fraction, hence the effect is much
more pronounced for the case $\delta = 0.1$ than for the
case $\delta = 2.0$. 

Another way of expressing the vast
discrepancy in the monomer crowding of the two 
systems is to look at the monomer packing fraction 
$\eta_{\rm m}$.
As there are $\nu$ monomers per dendrimer, this quantity is
given by the expression
\begin{equation}
\eta_{\rm m} = \frac{\pi}{6}\nu\rho\tau^3\left(\frac{\sigma}{\tau}\right)^3. 
\label{etam:eq}
\end{equation} 
For both types of dendrimers, $\nu = 62$ and $\rho_{\rm max}\tau^3 = 0.605$.
Yet, the ratio $\sigma/\tau$ has the value 0.188 for $\delta = 0.1$
and $0.101$ for $\delta = 2.0$, see the last column of Table 
\ref{TABparameters}. Accordingly, at $\rho = \rho_{\rm max}$ we obtain
$\eta_{\rm m} = 0.13$ for $\delta = 0.1$ but $\eta_{\rm m} = 0.02$ for
$\delta = 2.0$. The soft dendrimers have a much lower monomer packing
fraction at $\rho_{*}$ than the stiffer ones, a result that can be
traced to the fact that their radius of gyration is larger.\footnote{This  
is characteristic for non-compact objects: for polymer chains,
e.g., one obtains $\eta_{\rm m} \sim R_g^{-4/3}$ at the overlap
concentration \cite{likos:pr:01}.} Thus, we conclude that the 
density-dependence of the pair interaction can be traced back to
the shrinking of the dendrimers, a phenomenon that leads to increased
crowding of the monomers in their interior. 

\section{Total scattering intensities and the factorisation approximation}
\label{scatter:sec}

In this section we turn our attention to a different
question, which is however related to the issues discussed above,
namely to the interpretation of scattering data from concentrated
dendrimer solutions. As a first step, we consider the form factor
$F(q)$, defined by Eq.\ (\ref{fofq:eq}). Clearly, $F(q)$ expresses
the {\it intramolecular} correlations between the monomers 
belonging to a certain dendrimer. When scattering from an
infinitely dilute solution, $F(q)$ offers the only contribution
to the coherent scattering density. Since all the information
about the monomer correlations is encoded in $F(q)$, great
experimental effort is devoted to the determination of this
quantity. At low values of $q$, $qR_{g,\infty}$, the form
factor delivers information about the overall size of the molecule
whereas at higher values of the scattering wavevector, $q \sim 1/a$,
where $a$ is the monomer length, information about the 
monomer correlations and the fractal dimension of the object
is hidden \cite{likos:pr:01, benoit, grest:review}.

\begin{figure}
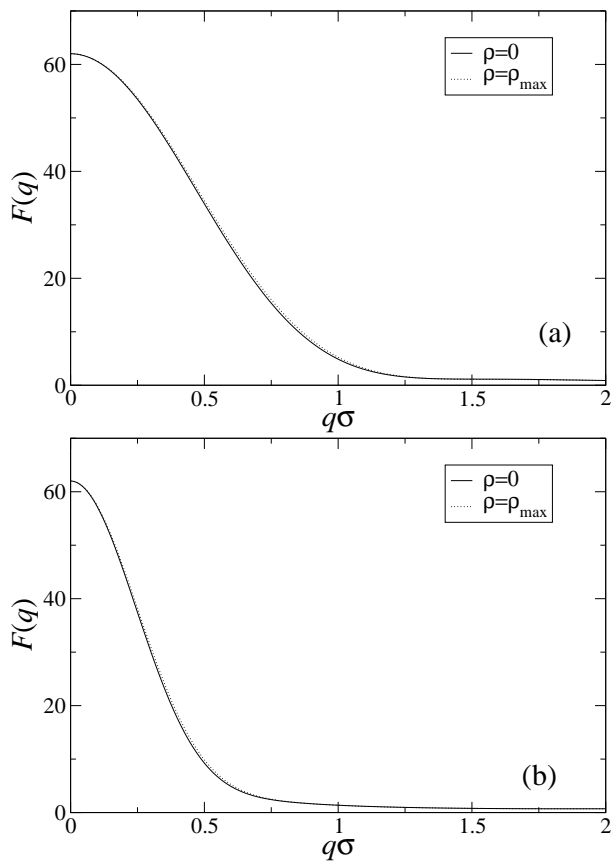

  \begin{center}
  \begin{minipage}[h]{7.2cm}
  \includegraphics[width=8cm,clip]{figure07a.eps}
  \includegraphics[width=8cm,clip]{figure07b.eps}
  \end{minipage}
  \end{center}
  \caption{The form factors measured in the monomer-resolved simulations
   for one isolated dendrimer molecule ($\rho = 0$, solid line) and
   at the highest density ($\rho = \rho_{\rm max}$, dotted line).
   The model dendrimers have 
   maximum thread length (a) $\delta=0.1$ and (b) $\delta=2.0$.}
  \label{G4_Fq}
\end{figure}

Although $F(q)$ is experimentally measured at the limit $\rho \to 0$,
the same quantity can be defined at any density. At arbitrary
concentrations, $F(q)$ will in general change with respect to its
form at infinite dilution, due to possible deformations of the
macromolecules. In Fig.\ \ref{G4_Fq} we show the form factors
for the two model dendrimers at the lowest and at the highest
simulate densities. It can be seen there that there is only 
a small change in both cases, which takes the form of a slight
extension of $F(q)$ to higher $q$-values as the concentration 
increases. This is consistent with the shrinkage of the dendrimers
and the corresponding decrease of the gyration radius. Indeed,
in the Guinier regime, $qR_g < 1$, the form factor has a parabolic
profile, $F(q) \cong N[1 - (qR_g)^2/3]$, and a reduction of
$R_g$ manifests itself as a swelling in $q$-space and vice versa
\cite{likos:pre:98}.

\begin{figure}
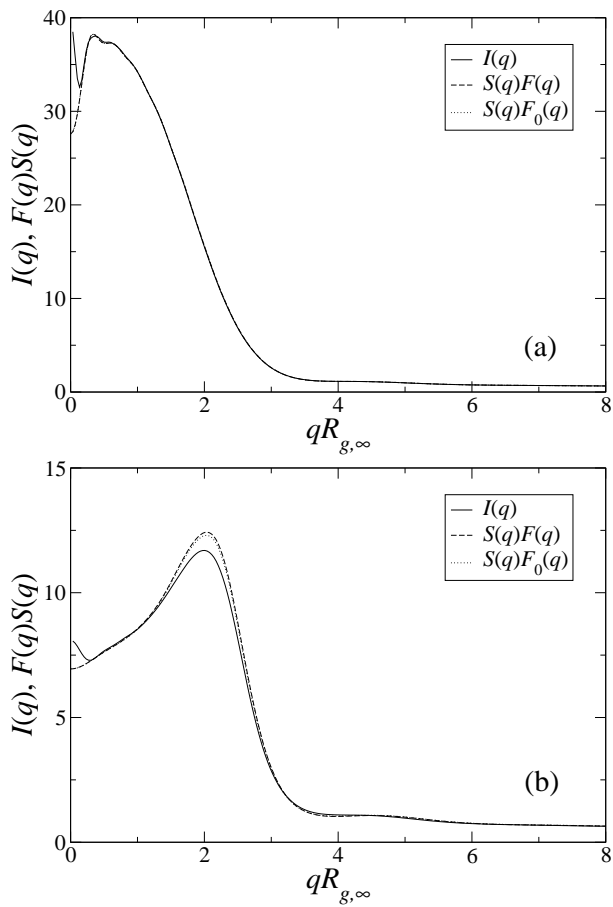

  \begin{center}
  \begin{minipage}[h]{7.2cm}
  \includegraphics[width=8cm,clip]{figure08a.eps}
  \includegraphics[width=8cm,clip]{figure08b.eps}
  \end{minipage}
  \end{center}
  \caption{The total coherent scattering intensity $I(q)$ 
           [Eq.\ (\ref{iofq:eq})] from concentrated 
           $\delta = 0.1$-dendrimer solutions, compared with the result from
           the factorisation approximation, Eq.\ (\ref{appr3:eq}),
           at different overall concentrations $\rho$.
           (a) $\rho = 0.1\rho_{\rm max}$ and (b)
               $\rho = 0.5\rho_{\rm max}$. Results using both
           the form factor $F(q)$ at the given density and its
           counterpart at infinite dilution, $F_0(q)$ are shown
           for the factorisation approximation.}
\label{fact1_0.1:fig}
\end{figure}

Let us now turn our attention to the total coherent scattering 
intensity from all monomers, $I(q)$, given by Eq.\ (\ref{iofq:eq}).
It is clear from its definition that $I(q)$ can also be measured
in the monomer-resolved simulation and this has been done for both
dendrimer species, characterised by the maximum thread extensions
$\delta = 0.1$ and $\delta = 2.0$.
In attempting to model complex polymeric entities as soft colloids,
it is a common procedure to separate the intramolecular from
the intermolecular correlations and to write down approximations
for the quantity $I(q)$ in which the two types of correlations
appear in a factorised fashion. Here we are going to put this
approach into a test and figure out the limits of its validity
as far as dendritic molecules are concerned. A similar test
has been carried out by Krakoviak {\it et al.} \cite{krak:epl:02}
who compared results from the PRISM model for polymers with
simulations and with the factorisation ansatz.

\begin{figure}
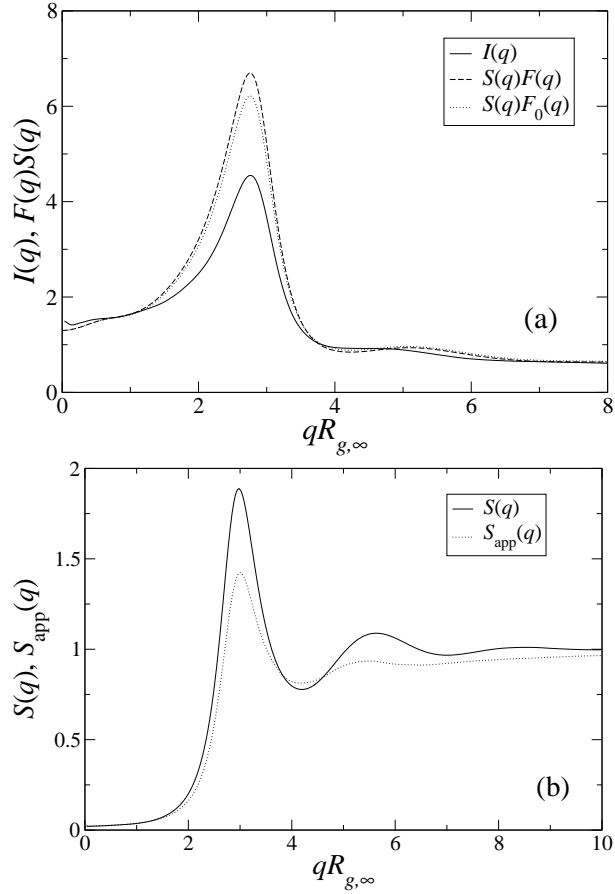

  \begin{center}
  \begin{minipage}[h]{7.2cm}
  \includegraphics[width=8cm,clip]{figure09a.eps}
  \includegraphics[width=8cm,clip]{figure09b.eps}
  \end{minipage}
  \end{center}
  \caption{(a) Same as Figs.\ \ref{fact1_0.1:fig}(a) and (b) but
  for $\rho = \rho_{\rm max}$. (b) The true structure factor
  $S(q)$ between the centres of mass at $\rho = \rho_{\rm max}$,
  as obtained from the monomer-resolved simulations, compared
  with the apparent structure factor $S_{\rm app}(q) = I(q)/F_0(q)$.}
  \label{fact2_0.1:fig}
\end{figure}

As a first approximate step, one assumes that the intramolecular
conformations and centre-of-mass correlations decouple from
each other. Correspondingly, Eq.\ (\ref{iofq:eq}) takes the 
approximate form:
\begin{equation}
\fl
I(q) \cong \frac{1}{N\nu}\sum_{i=1}^N\sum_{j=1}^N
                      \sum_{\alpha=1}^{\nu}\sum_{\beta=1}^{\nu}
                      \left\langle
                      \exp\left[-{\rm i}{\bf q}\cdot
                      \left({\bf r}_i - {\bf r}_j\right)\right]
                      \right\rangle
                      \left\langle
                      \exp\left[-{\rm i}{\bf q}\cdot
                      \left({\bf u}_{\alpha}^{i} 
                      - {\bf u}_{\beta}^{j}\right)\right]
                      \right\rangle.
\label{appr1:eq}
\end{equation}
The approximation inherent in Eq.\ (\ref{appr1:eq}) above is
a reasonable one for dendrimers. Indeed, as it has been shown
in Ref.\ \cite{harreis:jcp:03}, the monomer degrees of freedom
are correlated at length scales $\sim \sigma$, whereas for
the overall densities $\rho$ considered here, the centers
of mass are correlated at lengths at least $\sim R_g$ and the
two are well-separated from each other. Hence, at the 
wavevector-scale $q_{\rm CM} \sim 1/R_g$ at which 
the centre-of-mass $S(q)$ shows structure, the dendrimers
still appear as compact objects and the internal fluctuations
can be decoupled from the intermolecular ones. The second approximation
is now the following. Suppose that
we are at sufficiently low densities, so that close approaches
between the centres of mass of the dendrimers are very rare and
they carry therefore a negligible statistical weight. Then, since
monomers belonging to different dendrimers stay far apart, it
is reasonable to assume that the deviations from their respective
centres of mass are uncorrelated. In this case, one can approximately
write:
\begin{eqnarray}
\fl
\nonumber
\frac{1}{\nu}\sum_{\alpha=1}^{\nu}\sum_{\beta=1}^{\nu} 
                      \left\langle
                      \exp\left[-{\rm i}{\bf q}\cdot
                      \left({\bf u}_{\alpha}^{i} 
                      - {\bf u}_{\beta}^{j}\right)\right]
                      \right\rangle 
& \cong &
\frac{1}{\nu}\sum_{\alpha=1}^{\nu}\sum_{\beta=1}^{\nu} 
                      \left\langle
                      \exp\left(-{\rm i}{\bf q}\cdot
                      {\bf u}_{\alpha}^{i}\right)\right\rangle
                      \left\langle\exp\left({\rm i}{\bf q}\cdot 
                      {\bf u}_{\beta}^{j}\right)
                      \right\rangle 
\\
& = & \frac{1}{\nu}\left\langle
                   \hat\xi_{\bf q}\right\rangle
                   \left\langle\hat\xi_{-\bf q}
                   \right\rangle, 
\label{appr2:eq}
\end{eqnarray}
where $\xi_{\bf q}$ is the Fourier transform of the monomer density
operator $\hat\xi({\bf u})$ around the centre of mass of an arbitrary
dendrimer:\footnote{The quantity $\xi({\bf u})$ defined in
Eq.\ (\ref{xiofr:eq}) is simply the expectation value of the
operator $\hat\xi({\bf u})$.}
\begin{equation}
\hat\xi({\bf u}) = \sum_{\alpha = 1}^{\nu}\delta\left(
                                      {\bf u}-{\bf u}_{\alpha}^{i}\right).
\label{hatxi:eq}
\end{equation}
Clearly, the right hand side of Eq.\ (\ref{appr2:eq}) has no dependence
on the dendrimer index. At the same time, it has been shown in
Ref.\ \cite{harreis:jcp:03} that the product
$\nu^{-1}\langle\hat\xi_{\bf q}\rangle\langle\hat\xi_{-{\bf q}}\rangle$
is an excellent
approximation for the form factor $F(q)$ of the dendrimers, deviations
from the exact expression in Eq.\ (\ref{fofq:eq}),
$F(q) = \nu^{-1}\langle\hat\xi_{\bf q}\hat\xi_{-{\bf q}}\rangle$,
appearing only
at high $q$-values that are unreachable in a typical SANS experiment.
The approximation inherent in Eq.\ (\ref{appr2:eq}) has been derived
for monomers belonging to different dendrimers ($i \ne j$) and now,
in view of the results of Ref.\ \cite{harreis:jcp:03}, it can be also
applied to the case $i = j$. Putting everything together,
we obtain
\begin{equation}
\frac{1}{\nu}\sum_{\alpha=1}^{\nu}\sum_{\beta=1}^{\nu} 
                      \left\langle
                      \exp\left[-{\rm i}{\bf q}\cdot
                      \left({\bf u}_{\alpha}^{i} 
                      - {\bf u}_{\beta}^{j}\right)\right]
                      \right\rangle \cong F(q).
\label{rigid:eq}
\end{equation}
Eqs.\ (\ref{appr1:eq}) and (\ref{rigid:eq}) now yield
the oft-employed {\it factorisation approximation}:
\begin{equation}
I(q) \cong S(q)F(q),
\label{appr3:eq}
\end{equation}
whose validity will be tested in what follows. 

\begin{figure}
  \begin{center}
  \begin{minipage}[h]{7.2cm}
  \includegraphics[width=8cm,clip]{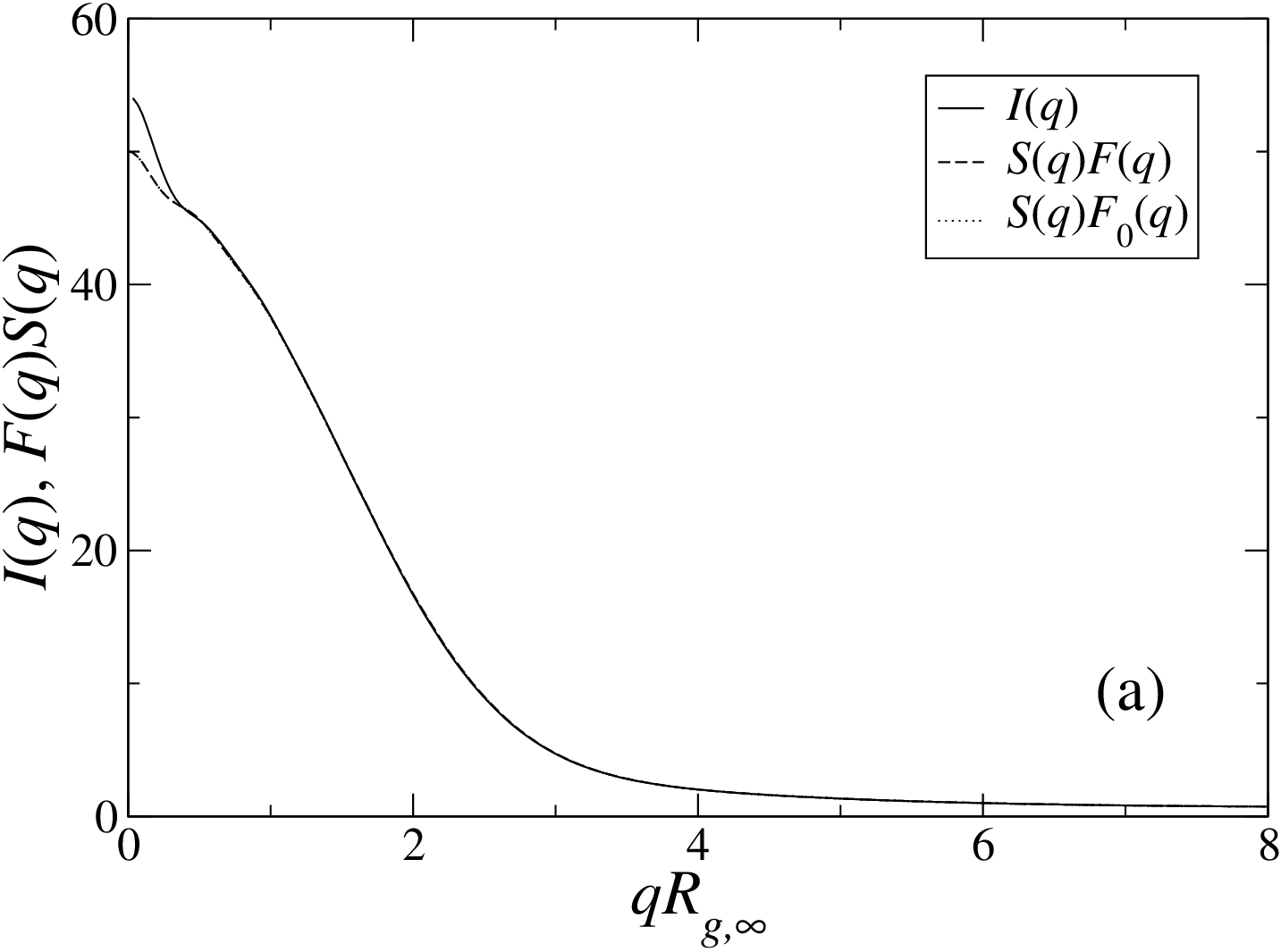}
  \includegraphics[width=8cm,clip]{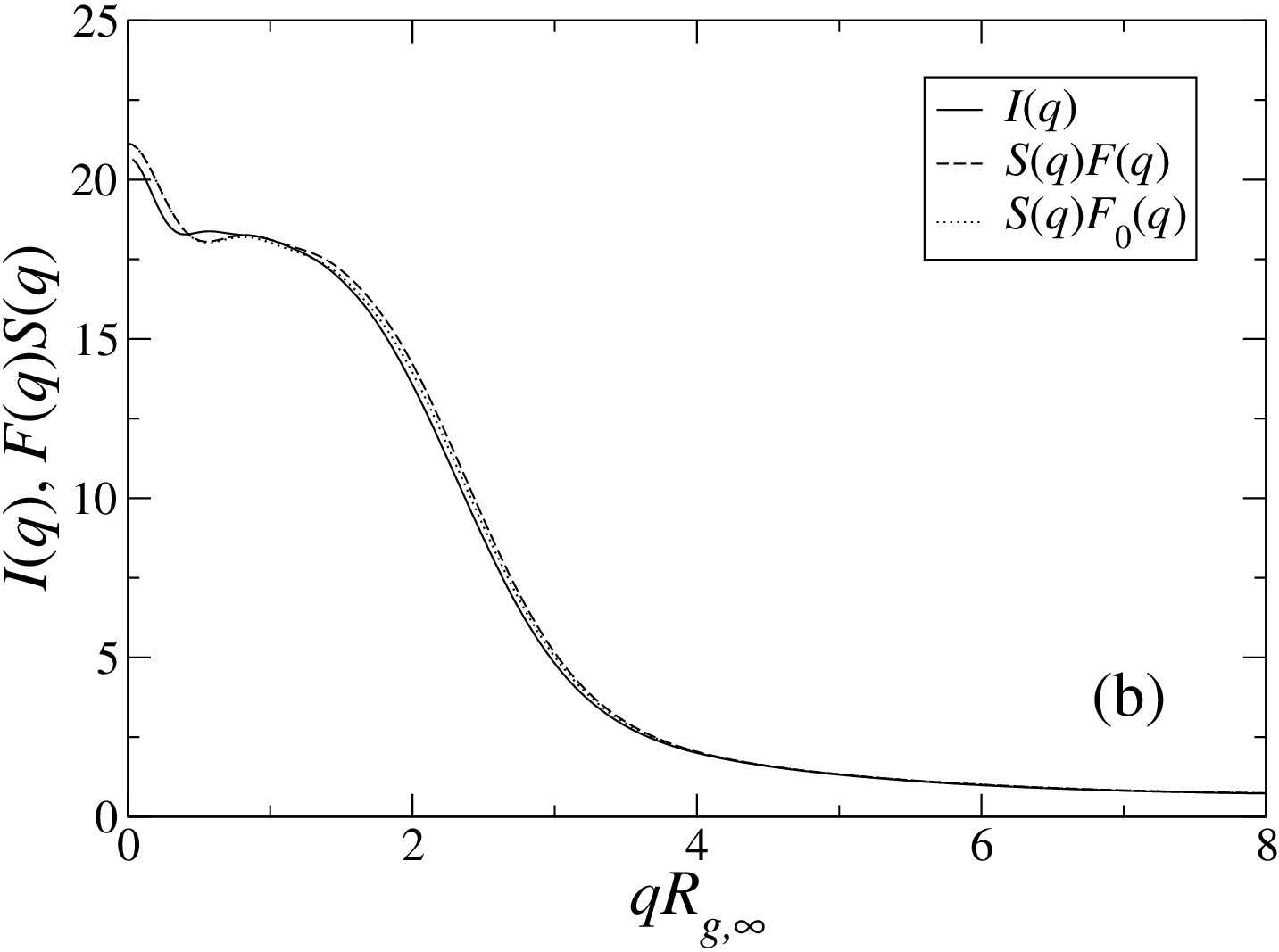}
  \end{minipage}
  \end{center}
  \caption{Same as Fig.\ \ref{fact1_0.1:fig} 
           but for $\delta = 2.0$-dendrimers.}
  \label{fact1_2.0:fig}
\end{figure}

\begin{figure}
  \begin{center}
  \begin{minipage}[h]{7.2cm}
  \includegraphics[width=8cm,clip]{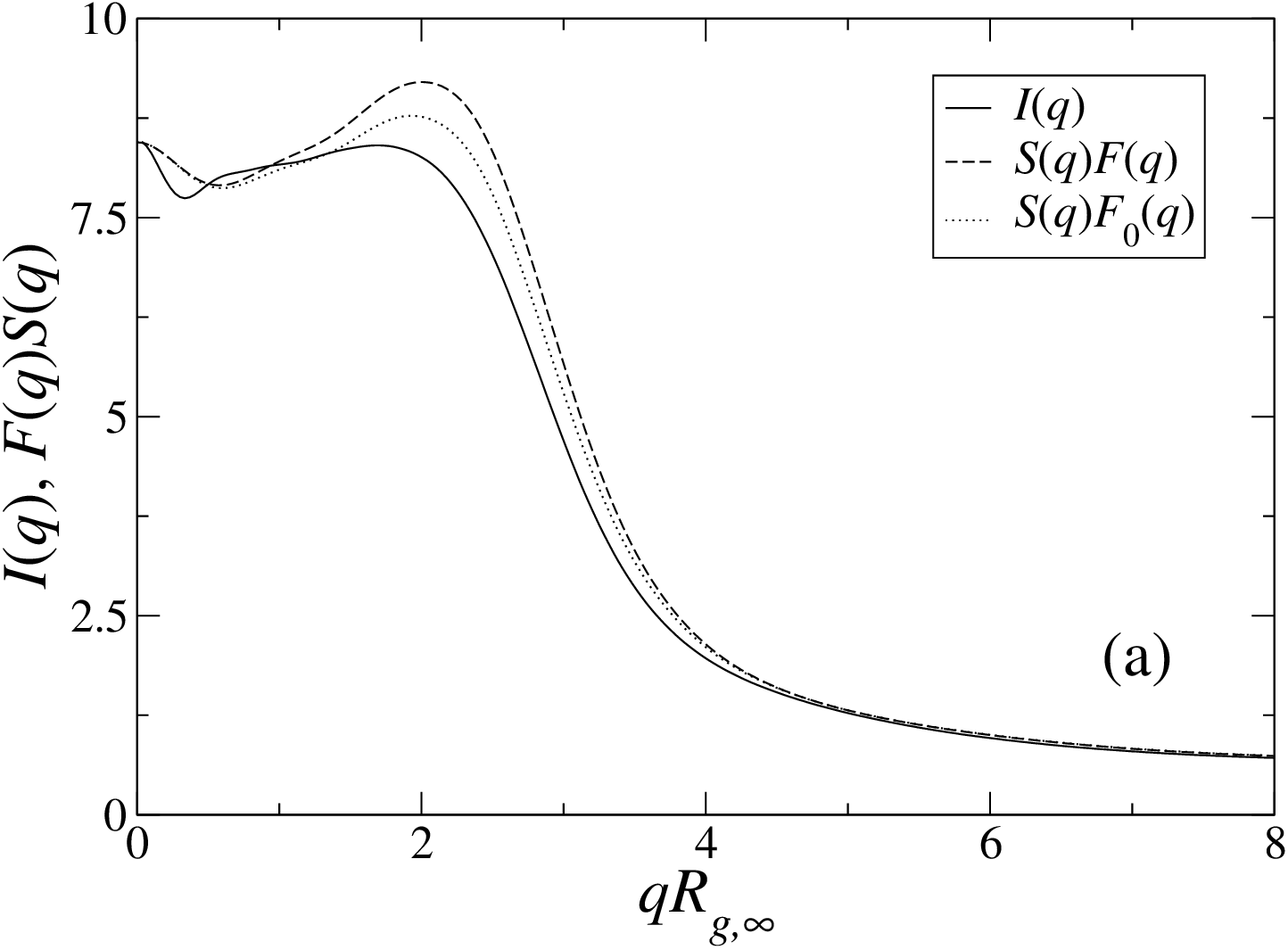}
  \includegraphics[width=8cm,clip]{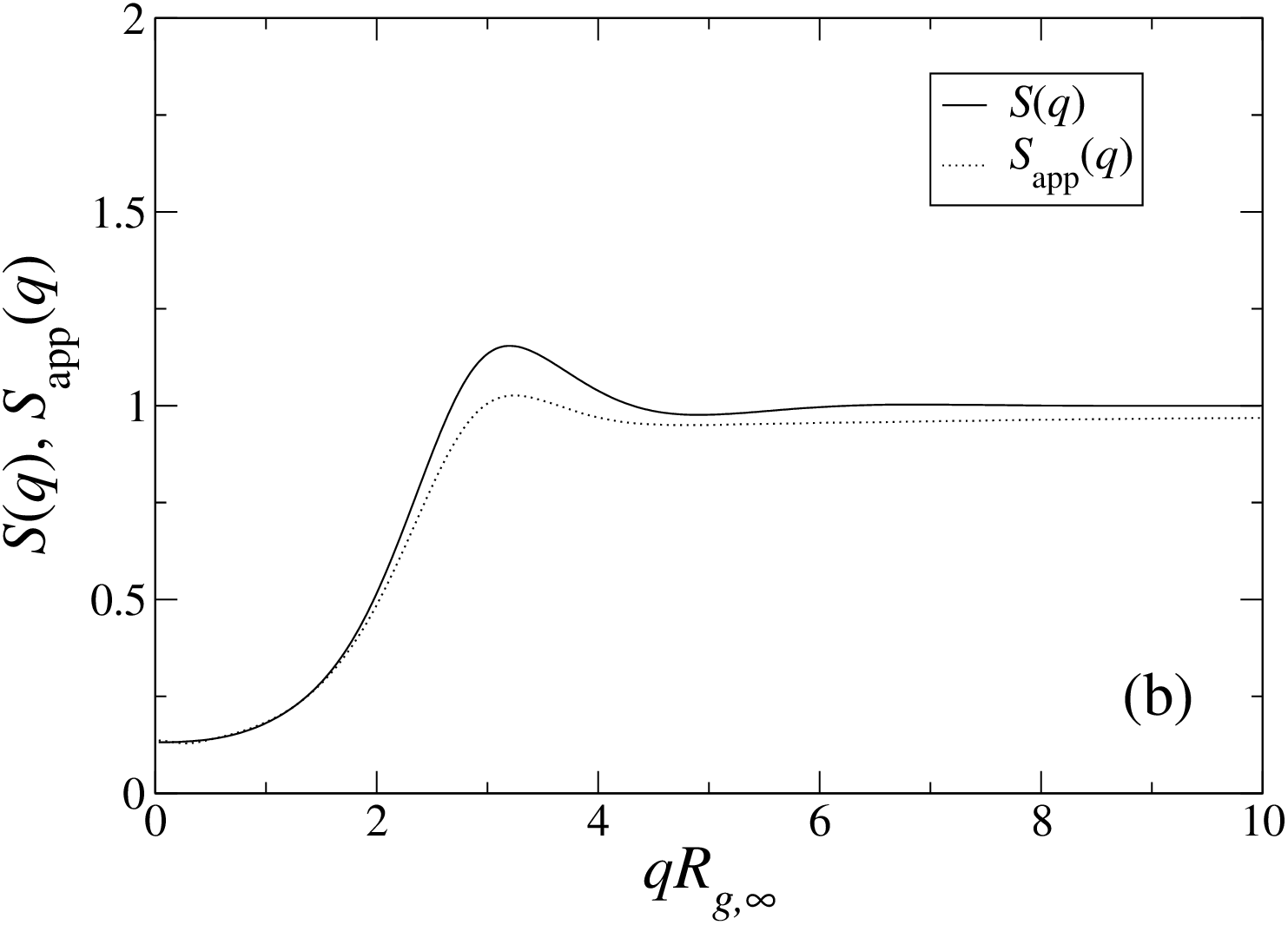}
  \end{minipage}
  \end{center}
  \caption{Same as Fig.\ \ref{fact2_0.1:fig} 
           but for $\delta = 2.0$-dendrimers.}
  \label{fact2_2.0:fig}
\end{figure}

The assumptions that went into the derivation of Eq.\ (\ref{appr3:eq})
above become exact when the particles from which one scatters are
rigid colloids \cite{klein:96}, in which case individual scattering centres
are devoid of a fluctuating nature. In this context, it is important to note
that there is an analog of the factorisation approximation that
is applied in the theory of concentrated polymer solutions
and carries the name ``rigid particle assumption'' 
\cite{krak:epl:02, cates:epl:01}. Here, one starts from 
Eq.\ (\ref{appr1:eq}) and assumes that 
monomer-monomer
correlations between monomers belonging to different polymers
are identical to the intramolecular correlations in any
chain \cite{krak:epl:02}.
Under this assumption, the second factor on the
right-hand-side of Eq.\ (\ref{appr1:eq}) above takes the form:
\begin{equation}
\fl
\frac{1}{\nu}\sum_{\alpha=1}^{\nu}\sum_{\beta=1}^{\nu} 
                      \left\langle
                      \exp\left[-{\rm i}{\bf q}\cdot
                      \left({\bf u}_{\alpha}^{i} 
                      - {\bf u}_{\beta}^{j}\right)\right]
                      \right\rangle \cong
\frac{1}{\nu}\sum_{\alpha=1}^{\nu}\sum_{\beta=1}^{\nu} 
                      \left\langle
                      \exp\left[-{\rm i}{\bf q}\cdot
                      \left({\bf u}_{\alpha}^{i} 
                      - {\bf u}_{\beta}^{i}\right)\right]
                      \right\rangle = F(q),
\label{cates:eq}
\end{equation}
and, in conjunction with Eq.\ (\ref{appr1:eq}), the factorisation approximation
of Eq.\ (\ref{appr3:eq}) follows once again. Krakoviak {\it et al.} tested
the validity of Eq.\ (\ref{appr3:eq}) for polymer solutions, finding that
it breaks down for high polymer densities.

We have put the validity of Eq.\ (\ref{appr3:eq}) into a strong test
by comparing the directly measured total coherent scattering intensity
$I(q)$ with the product $F(q)S(q)$, where for the latter quantity both
factors are the ones measured in the same simulation. Results are shown
in Figs.\ \ref{fact1_0.1:fig} and \ref{fact2_0.1:fig}(a) for the
$\delta = 0.1$-dendrimers as well as in Figs.\ \ref{fact1_2.0:fig} 
and \ref{fact2_2.0:fig}(a) for the $\delta = 2.0$-dendrimers. It can
be seen that the factorisation approximation is valid at the lowest
density shown ($\rho = 0.1\rho_{\rm max}$) but that its quality
becomes poorer as the concentration of the solution increases. 
A dramatic breakdown can be seen in Fig.\ \ref{fact2_0.1:fig}(a) for
the more compact dendrimers, whereas the breakdown is also clear (but less
spectacular) for the more open dendrimers, Fig.\ \ref{fact2_2.0:fig}(a).

We can now trace back to the physical origins of the breakdown of
the factorisation approximation, Eq.\ (\ref{appr3:eq}). There is
first of all a weak breakdown of the first assumption, Eq.\ (\ref{appr1:eq}),
in which the centre-of-mass coordinates were decoupled from the
fluctuating monomers. Indeed, were this approximation to be true,
then the form factor $F(q)$ would remain unchanged at all concentrations.
This is however not the case, as the results in Fig.\ \ref{G4_Fq}
demonstrate: the dendrimers shrink as $\rho$ grows. Yet, the difference
between the infinite-dilution form factor, $F_0(q)$ and its counterpart
at finite density, $F(q)$, is not sufficient to account for the
failure of the factorisation approximation. As can be seen in 
Figs.\ \ref{fact1_0.1:fig}(b), \ref{fact2_0.1:fig}(a),
\ref{fact1_2.0:fig}(b) and \ref{fact1_2.0:fig}(a), the product
$S(q)F(q)$ is in even {\it worse} agreement with $I(q)$ than
the product $S(q)F_0(q)$. The reason for the breakdown of 
Eq.\ (\ref{appr3:eq}) lies in the assumption inherent in 
deriving the approximation of Eq.\ (\ref{rigid:eq}), namely
that fluctuations between monomers belonging to different
dendrimers are uncorrelated. At sufficiently low densities
$\rho$, this is a reasonable assumption. However, in approaching
the overlap density $\rho_{*}$, it does not hold any more. As
monomers from different dendrimers begin to crowd with one another,
their coordinates with respect to their centers of mass become
more and more strongly correlated and Eq.\ (\ref{rigid:eq}) loses
its validity. In this respect, it is not surprising that the
breakdown of Eq.\ (\ref{appr3:eq}) is more dramatic for the
$\delta = 0.1$-dendrimers than for the $\delta = 2.0$-ones. 
In the former case, the monomer packing fraction is higher and
the corresponding correlations between monomers belonging to 
different molecules stronger than in the latter. To put it
in more pictorial terms: at the overlap concentration it is
not any more possible to tell to which dendrimer a monomer
belongs, see Fig.\ \ref{rho10:fig}. A clear separation between
intra- and inter-dendrimer fluctuations is not any more possible.

We finally discuss the consequences of the above findings for
the interpretation of scattering data obtained from concentrated
dendrimer solutions. The validity of Eq.\ (\ref{appr3:eq}) is
often taken for granted: the form factor $F(q)$ is measured
in a SANS- or SAXS experiment at low concentrations and
extrapolated to infinite dilution to obtain the quantity
$F_0(q)$. Thereafter, the measured coherent scattering intensity at
any concentration, $I(q)$ is divided through $F_0(q)$, the result
being interpreted as the structure factor of the system. 
In order to differentiate it from $S(q)$, we emphasise here
that this is only an {\it apparent} structure factor $S_{\rm app}(q)$,
given by
\begin{equation}
S_{\rm app}(q) = \frac{I(q)}{F_0(q)}.
\label{sapp:eq}
\end{equation} 
In Figs.\ \ref{fact2_0.1:fig}(b) and \ref{fact2_2.0:fig}(b) we
compare the apparent structure factors for the two dendrimer species
at the highest simulated density with the true ones. It can be
seen that the process of applying Eq.\ (\ref{sapp:eq}) has the 
effect of producing apparent structure factors that are everywhere
lower than the true ones and they even fail to reach the asymptotic
value unity at the range considered. 

Such structure factors
from concentrated dendrimer solutions have been published in
Refs.\ \cite{topp:macrom:99} and \cite{ramzi:macrom:98}, 
in which they have been correctly
termed `apparent'. It is important here to point out that 
apparent structure factors can lead to false conclusions
regarding the validity of the pair potential approximation 
in mesoscopic theories of dendrimer solutions. Indeed, as we
have explicitly shown in this work, many-body effective potentials play
only a minor role in concentrated dendrimer solutions, therefore,
one can obtain accurate structure factors from theory by working
with a density-independent pair potential. If, however, these
structure factors were to be compared with the apparent experimental
quantities $S_{\rm app}(q)$, discrepancies of the kind shown in
Figs.\ \ref{fact2_0.1:fig}(b) and \ref{fact2_2.0:fig}(b) would
show up. It would be then possible to argue that these discrepancies
are due to the breakdown of the pair potential approximation but,
as we have shown here, this conclusion would be unwarranted. 
The reason for the disagreement between
theory and `experiment' would, in this case, lie in the
employment of an {\it erroneous} approximation, Eq.\ (\ref{appr3:eq}),
in deriving apparent structure factors from the experimental data.
It is worth noting that Krakoviak {\it et al.} \cite{krak:epl:02}
reached similar conclusions for the case of polymer solutions, although
they did not formally introduce an apparent structure factor into
their considerations.  

\section{Summary and concluding remarks}
\label{summary:sec}

We have carried out extensive, monomer-resolved and effective
simulations of model dendrimers in order to calculate 
correlation functions between the centres of mass of the
macromolecules and the individual monomers themselves.
By comparing the real-space correlation functions obtained
by the two simulation approaches, we found that many-body
effective potentials play a minor role up to the overlap 
density and they can be altogether ignored for open dendrimers
with long bond lengths. Our finding for the scattering intensity,
on the other hand, is that the factorisation approximation of
this quantity into a form- and a structure factor loses its
validity as one approaches the overlap concentration. 
Structure factors that are obtained from experimental data
by dividing the scattering intensity through the form factor
can be seriously in error. 

It appears, therefore, that the extraction of an accurate structure
factor from concentrated dendrimer solutions is extremely difficult as
one approaches the overlap concentration. We anticipate that this
result is also valid for other `polymeric colloids' such as
star-shaped polymers and brushes. One strategy to circumvent this
inherent difficulty is to use the labeling technique, in which
a small, inner part of the molecule is protonated and the rest
is deuterated in such a way that the contrast between the outermost
part of the molecule and the solvent vanishes. In this way, only
the innermost part of the molecule will have contrast with
the solvent and scatter coherently. Thus, one can reach concentrations
for the whole system that exceed $\rho_{*}$, whereas the labeled
parts are still nonoverlapping. Such a technique was
successfully applied, e.g.,
to star polymers \cite{likos:prl:98}.

\section*{Acknowledgments}
This work has been supported by the Deutsche Forschungsgemeinschaft (DFG).


\begin{thebibliography}{99}
\bibitem{foot:charge} It is intriguing, in this respect, that the three-body
forces are also attractive for charged colloids, see Refs.\ \cite{dobn:prl:04}
and \cite{dobn:pre:04}. However, in the latter case the many-body forces
arise through nonlinear counterion screening and the corresponding
rearrangements of the counterion clouds, hence a direct analogy with
the case at hand cannot be made.
\end{thebibliography}
\end{document}